\documentclass[
    10pt
    ,aps
    ,pra
    ,oneside
    ,nobibnotes
    ,floatfix
    ,twocolumn
    ,nofootinbib
    ,superscriptaddress
    ,showkeys
    ,fleqn
]{revtex4-2}

\usepackage[english,russian]{babel}
\providecommand{\English}{\selectlanguage{english}\color{black}}

\usepackage{mathtools}
\usepackage{array}[=2016-10-06]

\usepackage{tabularx}

\usepackage{mleftright}

\usepackage{derivative}

\usepackage{graphicx}
\graphicspath{{}{Graph/}{Graph/seminar/}{Graph/PNG/}
    {../Graph/}{../Graph/A0}{../Graph/A1}{../Graph/A2}{../Graph/A3}{../Graph/GF}
    {../../Graph/}{../../Graph/A0}{../../Graph/A1}{../../Graph/A2}{../../Graph/A3}{../../Graph/GF}
}

\usepackage{siunitx}
\NewDocumentCommand{\bnum}{O{}m}{%
  \textbf{\num[color=Blue,#1]{#2}}%
}
\DeclareSIUnit{\gauss}{G}

\usepackage[svgnames]{xcolor}

\usepackage[normalem]{ulem}
\usepackage{silence}

\let\textcite\relax
\let\citet\relax

\let\citep\relax

\expandafter\let\csname ver@natbib.sty\endcsname\relax
\expandafter\let\csname write@bibliographystyle\endcsname\relax
\usepackage[%
 language=auto,
 autolang=other*,
 natbib=true,
 bibstyle=numeric,
 citestyle=numeric-comp,
 sorting=none,
]
{biblatex}
\usepackage[
    colorlinks
        ,linkcolor=violet
        ,filecolor=purple
        ,citecolor=teal
        ,urlcolor=magenta
    ,unicode
    ,pdfpagemode=UseOutlines
    ,pdfdisplaydoctitle=true
    ,pdfpagetransition=Wipe
    ,pdfusetitle
]{hyperref}
\hypersetup{
    ,pdfkeywords={GDMT, LoDestro equation, ballooning modes}
}
\usepackage{bookmark}

\iftutex
\usepackage{fontspec}

\usepackage[math-style=ISO,bold-style=ISO]{unicode-math}
\else

\usepackage[utf8]{inputenc} 

\usepackage{cmap}
\fi

    \iftutex
        \renewcommand{\vec}[1]{\symbf{#1}}
    \else
        \renewcommand{\vec}[1]{\mathbf{#1}}
    \fi

    \DeclareMathOperator{\vnabla}{\vec{\nabla}}

    \DeclareMathOperator{\const}{const}

\makeatletter
\@ifpackageloaded{unicode-math}{%
  \providecommand{\tensor}[1]{\mathbfsfit{#1}}%
  \PackageInfo{handy}{\string\tensor\space has been declared using ^^J\space\space \string\overleftrightarrow\space from unicode-math package}%
}{%
  \@ifpackageloaded{accents}{%
	\providecommand{\tensor}[1]{\accentset{\leftrightarrow}{{\symbf{#1}}}}%
    \PackageInfo{handy}{\string\tensor\space has been declared using ^^J\space\space \string\accentset\space from in accents package}%
  }{%
    \@ifpackageloaded{revsymb4-1}{%
      \PackageWarningNoLine{handy2}{\string\tensor\space has been taken from revsymb4-1 package.^^J You are recommended to load `accents' package before me}%
    }{
    \@ifpackageloaded{revsymb}{%
        \PackageWarningNoLine{handy2}{\string\tensor\space has been taken from revsymb package.^^J You are recommended to load `accents' package before me}%
      }{%
        \ifx\undefined\overset
        \else
          \providecommand{\tensor}[1]{\overset{\leftrightarrow}{\vec{#1}}}
          \PackageWarningNoLine{handy2}{\string\tensor\space has been declared using ^^J\space\space \string\overset\space from amsmath package.^^J You are recommended to load `accents' package before me}%
        \fi
      }%
    }%
  }%
}%
\makeatother
\usepackage{xr-hyper}
\usepackage[
    colorlinks
        ,linkcolor=violet
        ,filecolor=purple
        ,citecolor=teal
        ,urlcolor=magenta
    ,unicode
    ,pdfpagemode=UseOutlines
    ,pdfdisplaydoctitle=true
    ,pdfpagetransition=Wipe
    ,pdfusetitle
]{hyperref}
    \hypersetup{
        ,pdfkeywords={GDMT, non-paraxial plasma equilibrium, ballooning modes}
    }
\usepackage{bookmark}
\usepackage{booktabs}

\hypersetup{
    ,pdfkeywords={ballooning modes, boundary conditions, the LoDestro equation}
}

\begin{document}
\English

\title{Mercier--Cotsaftis and Grad--Shafranov equations for anisotropic plasma}
\author{Igor Kotelnikov}

\begin{abstract}
  
  In this brief review, the historical aspects of the generalization of the Grad--Shafranov equation to the case of anisotropic plasma are discussed.
\end{abstract}
\date{\today}

\maketitle

\section{Preface}
\label{s01}

%
In a recent paper, Mikhail Khristo and Alexey Beklemishev \cite{KhristoBeklemishev2025JPP_91_E3} explore plasma equilibrium in a so-called diamagnetic trap \cite{Beklemishev2016PoP_23_082506}. Diamagnetic trap differs from traditional mirror traps in that it almost completely expels the magnetic field from the plasma volume, causing conventional magnetohydrodynamic equations to fail. These authors base their numerical simulation of equilibrium in a diamagnetic trap on equation (3.4) in \cite{KhristoBeklemishev2025JPP_91_E3}, which they write in the form
\begin{gather}
\label{00:00}
r\partial_{r}\left(
    r^{-1}\partial_{r}\psi
\right)
+
\partial^{2}_{z}
=
-
8\pi^{2}c^{-1}r\,J_{\theta}
\end{gather}
%
and call it the Grad--Shafranov equation.\footnote{
    In Ivan Chernoshtanov's preprint \cite{Chernoshtanov2025arXiv2512_01780}, a similar equation (4) is also called the Grad--Shafranov equation.
}
There is indeed some similarity with the Grad--Shafranov equation, since the operator on its left-hand side is called the Shafranov operator. However, in the canonical Grad--Shafranov equation (GSE), the right-hand side is a function of the magnetic flux $\psi(r,z)$, but not of its derivatives. The equation \eqref{00:00} obviously does not have this property, since in a diamagnetic trap the diameter of the fast ion orbit is approximately equal to the diameter of the plasma column and, consequently, the relationship of the diamagnetic current $J_{\theta}$ with the function $\psi(r,z)$ is nonlocal. This means that the equation \eqref{00:00} is integro-differential, whereas the Grad--Shafranov equation belongs to the class of nonlinear partial differential equations (PDE).


\begin{table*}[tb!]
\parbox[t]{0.6\linewidth}{
    \setlength{\extratabsurround}{8pt}
    \setlength{\extrarowheight}{4pt}
    \caption{Summary statistics of the number of publications}
    \label{tab:google}
    \begin{tabular}{l}
    \begin{tabular}{|l|c|c|c|c|}
    \hline
    & \multicolumn{2}{c|}{{Grad--Shafranov equation}} & \multicolumn{2}{c|}{{Mercier--Cotsaftis equation}} \\ \cline{2-5}
    {Publication} & \textit{Narrow term} & \textit{+ analogs*} & \textit{Narrow term} & \textit{+ context**} \\ \hline
    Articles & 800\ldots 1200 & 2500\ldots 3500 & 40\ldots 70 & 150\ldots 250 \\ \hline
    Monographs & 15\ldots 30 & 60\ldots 80 & 5\ldots 10 & 15\ldots 20 \\ \hline
    Textbooks & 40\ldots 60 & 120\ldots 150 & 10\ldots 15 & 20\ldots 25 \\ \hline
    Google Scholar & 1,100 & 4,800 & 50 & 350 \\ \hline
    \end{tabular}
    \\[12pt]
    *Including modified, anisotropic, relativistic, and flux GSE. 
    \\
    **Including papers that use MCE for deriving the Mercier criterion.
    \end{tabular}
}
\parbox[t]{0.38\linewidth}{
    \setlength{\extratabsurround}{8pt}
    \setlength{\extrarowheight}{4pt}
    \centering
    \caption{Term mentions in leading journals}
    \label{tab:journals}
    \begin{tabular}{|l|c|c|c|c|c|}
    \hline
    {Search query} & {PoP} & {JPP} & {NF} & {PPCF} & {Total} \\ \hline
    Generalized GSE & 112 & 48 & 32 & 41 & 233 \\ \hline
    Modified GSE & 142 & 44 & 37 & 51 & 274 \\ \hline
    Anisotropic GSE & 68 & 23 & 19 & 26 & 136 \\ \hline
    GSE with flow & 184 & 58 & 88 & 112 & 442 \\ \hline
    GSE + anisotropy & 315 & 108 & 142 & 158 & 723 \\ \hline
    MCE & 11 & 4 & 7 & 14 & 36 \\ \hline
    \end{tabular}
}
\end{table*}

%
Puzzled by this observation, I decided to examine the existing terminology, as using inappropriate equation names can make the problem difficult to understand. As a result, an entire stratum of publications and knowledge can be overlooked. This is how this mini-review of the known variations of the Grad--Shafranov equation in the scientific literature, as applied to plasma with anisotropic pressure, appeared. As it turns out, approximately the same form of a suitable equation has many names: the Generalized Grad--Shafranov equation (GGSE), the Modified Grad--Shafranov equation (MGSE), the Anisotropic Grad--Shafranov equation (AGSE), and, finally, the Mercier--Cotsaftis equation (MCE). Table \ref{tab:google} presents the results of an approximate count of the number of scientific articles, monographs, and textbooks that mention these names, with all variants mentioning the Grad--Shafranov equation combined into one column.

%
These estimates are very approximate due to the uncertainty of the counting criteria. For example, one may or may not include articles that don't explicitly mention the ``Mercier--Cotsaftis equation'' but actually use it to derive the Mercier stability criterion. Many authors don't use the word ``generalized'' in the title, simply calling their article ``Plasma Equilibrium \ldots ,'' but within the text, they re-derive the Generalized Grad--Shafranov equation. Therefore, the actual number of references to the GGSE/MGSE/AGSE/\ldots\ concept in texts is significantly higher (over 10,000).

%
The generalized Grad--Shafranov equation is a ``living'' tool of modern science, while the Mercier--Cotsaftis equation is more often regarded as a historical foundation or a special theoretical case. The generalized Grad--Shafranov equation is actively used in both Western and Russian schools of thought. The term ``Mercier--Cotsaftis equation'' is more specific to the French school of physics and is also encountered in Russian-language literature as a set phrase, although it has recently ``raised from the ashes.''

%
Since the term ``Mercier--Cotsaftis equation'' rarely varies linguistically (no one says ``modified Mercier--Cotsaftis equation''), it's best to search for related physical categories; for example, 90\% of the time the equation appears in the context of ``Mercier criterion derivation.'' Articles on the topic ``local stability of magnetic surfaces'' use this mathematical apparatus without mentioning Cotsaftis.
%
%
The approximate number of search results for specific phrases in leading plasma physics journals (PoP --- Physics of Plasmas, JPP --- Journal of Plasma Physics, NF --- Nuclear Fusion, PPCF --- Plasma Physics and Controlled Fusion) is presented in Table \ref{tab:journals}.

\section{Equilibrium equation for anisotropic plasma}
\label{s02}

%
The equation of Harold Grad \cite{GradRubin1958JNuclEnergy_7_284, GradRubin1958UN2_31_190} (1958) and Vitaly Shafranov \cite{Shafranov1957JETPh_33_710, Shafranov1958SovPhysJETP_6_545, Shafranov1960} (1957) is a fundamental tool for describing magnetohydrodynamic (MHD) equilibrium in axisymmetric toroidal systems. In the canonical derivation of the Grad--Shafranov equation (GSE), the plasma pressure is treated as an isotropic scalar $p(\psi)$, depending only on the magnetic flux $\psi$. However, with intense additional heating by injecting beams of neutral atoms (NBI) or radio frequency (RF) field, anisotropy arises, requiring the use of a pressure tensor:
\begin{gather}
\label{01:01}
\mathbf{P} 
= 
p_{\perp} \mathbf{I} + \left(p_{\|} - p_{\perp}\right) \vec{b}\vec{b}
,
\end{gather}
%
where $\vec{b} = \vec{B}/B$ is a unit vector along the magnetic field $\vec{B}$, and the functions $p_{\|}(\psi,B)$ and $p_{\perp}(\psi,B)$ simulate the plasma pressure across and along the magnetic field and, in addition to the magnetic flux, also depend on the magnitude of magnetic field $B=|\vec{B}|$.

%
The equilibrium of an axisymmetric plasma with a substantially anisotropic pressure ($p_{\|} \neq p_{\perp}$) is described by a generalization of the classical Grad--Shafranov equation (GSE). The mathematical foundation was laid in the works of Claude Mercier and Michel Cotsaftis \cite{MercierCotsaftis1961CompRendAcadSci_252_2203, MercierCotsaftis1961NF_1_121} (1961). Six years later, the Mercier--Cotsaftis equation (MCE) was reintroduced by Harold Grad \cite{Grad1967ProcSympApplMath_18_162, Grad1967PF_10_137} (1967), without mentioning the names of his predecessors.

\subsection{Canonical Grad--Shafranov equation}
\label{s1.01}

\begin{figure}
\includegraphics[width=\linewidth]{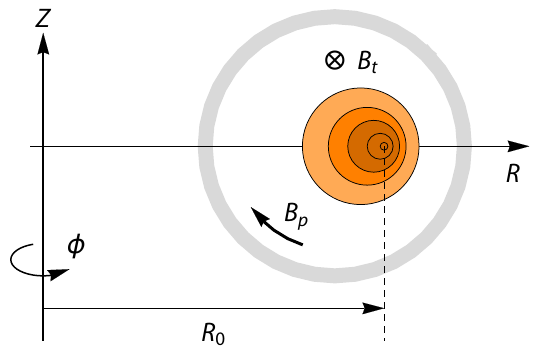}
\caption[Magnetic field in a symmetric tokamak]{
    The theory of axially symmetric tokamaks uses a cylindrical coordinate system $\left\{R,\phi,Z\right\}$, in which the symmetry axis $Z$ is directed along the major axis of the torus, and the coordinate $R$ is along the major radius of the torus. The components of the magnetic field in a symmetric tokamak can be expressed through the poloidal magnetic flux $\psi(R,Z)$ and the poloidal current $i_{p}(R,Z)$ through a circle of radius $R$ in the plane $Z=\const$ using the formulas \eqref{01.01:001}.
    In a symmetric tokamak, the magnetic field components $\vec{B}=\left\{B_{R},B_{\phi},B_{Z}\right\}$ do not depend on the toroidal angle $\phi$, as do the functions $\psi$ and $i_{p}$. The toroidal magnetic field $\vec{B}_{t}=\left\{0,B_{\phi},0\right\}$ is directed along the major circumference of the torus. The poloidal magnetic field $\vec{B}_{p}=\left\{B_{R},0,B_{Z}\right\}$ lies in the plane of section $\left\{R,Z\right\}$, perpendicular to the minor axis of the torus $R=R_{0}$, $Z=0$.
    The internal magnetic surfaces $\psi=\const$ of the minor axis of the torus are shifted toward the outer circumference of the torus more strongly than the external ones. This phenomenon is called the Shafranov shift.
}
\label{fig:209-01}
\end{figure}

Let us first recall the Grad--Shafranov equation (GSE) in its original, canonical form. It is used to describe the macroscopic equilibrium of isotropic plasma in axisymmetric magnetic traps (for example, in tokamaks) and is a nonlinear elliptic PDE for the poloidal magnetic flux function $\psi$. Axial symmetry means that in cylindrical coordinates $(R, \phi, Z)$ both the poloidal magnetic flux $\psi=\psi(R, Z)$ and the poloidal current flux $i_{p}=i_{p}(R, Z)$ do not depend on the azimuthal (toroidal) angle $\phi$, so the magnetic field and current density can be expressed in terms of $\psi(R, Z)$ and $i_{p}(R, Z)$ as follows:
\begin{align}
\label{01.01:001}
B_{R}   &= -\frac{1}{R}\pdv{\psi_{}}{Z}
,&
B_{\phi} &= \mu_{0}\frac{i_{p}}{R}
,&
B_{Z} &= \frac{1}{R}\pdv{\psi_{}}{R}
,\\
\label{01.01:002}                      
J_{R} &= -\frac{1}{R}\pdv{i_{p}}{Z}
,&
J_{\phi} &= - \frac{1}{\mu_{0}R}\, \Delta^{\ast}\psi_{}
,&
J_{Z} &= \frac{1}{R}\pdv{i_{p}}{R} ,
\end{align}
where $\mu_{0}=4\pi\times 10^{-7}\unit{\henry\per\metre}$ is the magnetic constant and
\begin{gather}
\label{01.01:003}
\Delta^* \psi
=
R \pdv{}{R}
\left( \frac{1}{R} \pdv{\psi}{R} \right)
+
\pdv[2]{\psi}{Z}
\end{gather}
is the Shafranov operator. Looking at Fig.~\ref{fig:209-01}, we can relate the reduced (i.e.\ divided by $2\pi$) magnetic flux to cylindrical coordinates using the integral
\begin{gather}
\label{01.01:004}
\psi(R,Z)
=
\int_{0}^{R}
R'\, B_{Z}(R',Z)\odif{R'}
\end{gather}
and similarly
\begin{gather}
\label{01.01:005}
i_{p}(R,Z)
=
\int_{0}^{R}
R' \,J_{Z}(R',Z)\odif{R'}
.
\end{gather}
In a vector form this reads
\begin{gather}
\label{01.01:006}
\vec{B} 
= 
\frac{1}{R} \left( \vnabla \psi \times \vec{e}_{\phi} \right) 
+ 
\mu_{0}\frac{i_{p}}{R} \vec{e}_{\phi}
,
\\
\label{01.01:007}
\vec{J} 
= 
\frac{1}{R} \left( \vnabla i_{p} \times \vec{e}_{\phi} \right) 
- 
\frac{1}{\mu_{0}R}\, \Delta^{\ast}\psi\, \vec{e}_{\phi}
,
\end{gather}
where $\vec{e}_{\phi}=\vnabla\phi/|\vnabla\phi|=R\,\vnabla\phi$ is the unit vector in the direction of the toroidal angle $\phi$. Under the given assumptions, the Grad--~Shafranov equation has the following form
\begin{gather}
\label{01.01:008}
\Delta^* \psi 
= 
-\mu_{0} R^{2} \odv{p}{\psi} - \mu_{0}^{2}\,i_{p} \odv{i_{p}}{\psi}
,
\end{gather}
%
where the functions $p(\psi)$ and $i_{p}(\psi)$ on the right-hand side are considered to be known functions of the magnetic flux.

%
The Grad--Shafranov equation is a consequence of the MHD equilibrium condition
\begin{gather}
\label{01.01:011}
\vnabla p = \vec{J} \times \vec{B}
\end{gather}
%
and Maxwell's equations. It expresses the balance of forces in which the plasma pressure gradient $\vnabla p$ is balanced by the Lorentz force $\vec{J} \times \vec{B}$. Solution of equation \eqref{01.01:003} allows us to determine the geometry of the magnetic surfaces $\psi(R,Z)=\const$ and the distribution of currents required for stable containment of the plasma column. The derivation of the canonical Grad--Shafranov equation is given in Appendix \ref{A1}. It is also proven there that in an isotropic plasma, the functions $p(\psi)$ and $i_{p}(\psi)$ do indeed depend on the coordinates $R$ and $Z$ only through the dependence of the magnetic flux $\psi=\psi(R,Z)$ on them. Such functions are called surface functions. 


Equation \eqref{01.01:006} first appeared under number (48) in Vitaly Shafranov's paper \cite{Shafranov1957JETPh_33_710}, published in the September 1957 issue of the leading Soviet physics journal JETP. The English-translated version  \cite{Shafranov1958SovPhysJETP_6_545} was published in March 1958. Harold Grad and Hanan Rubin reported the same equation in September 1957 at the Second United Nations Conference on the Peaceful Uses of Atomic Energy. In the thesis of their report \cite{GradRubin1958JNuclEnergy_7_284} this equation is given without a number, whereas in the conference proceedings \cite{GradRubin1958UN2_31_190} it has the number (18).

\medskip

Mathematical properties of the Grad--Shafranov equation have been studied in a number of papers and monographs. It has been shown that it can have zero, one, or more nontrivial solutions \cite{Kostomarov+2008MathModelling_20_1, HamFarrell2024NF_64_034001} depending on how the boundary value problem is posed.

\subsection{Generalized Grad--Shafranov equation}

\label{s01.02}

%
In the case of anisotropic pressure, typical for plasma with powerful additional heating, the pressure tensor in the so-called two-pressure model has the form \eqref{01:01}, so that the equilibrium condition 
\begin{gather}
\label{01.02:001}
\vnabla \cdot \mathbf{P} 
= 
\vec{J} \times \vec{B}
\end{gather}
for a stationary (non-rotating, non-flowing) plasma leads to the generalized Grad--Shafranov equation
\begin{gather}
\label{01.02:002}
\Delta^* \psi 
= 
- 
\frac{1}{\sigma} \vnabla \psi \cdot \vnabla \sigma
-
\frac{\mu_{0} R^{2}}{\sigma} \pdv{p_{\|}}{\psi} 
- 
\mu_{0}^{2}
\frac{F}{\sigma^{2}} \odv{F}{\psi} 
,
\end{gather}
%
where the magnetic anisotropy parameter
\begin{gather}
\label{01.02:003}
\sigma = 1 - \mu_{0} \frac{p_{\|} - p_{\perp}}{B^{2}}
\end{gather}
%
is introduced and modified current flux function is defined as
\begin{gather}
\label{01.02:005}
F
=
\sigma\, i_{p}.
\end{gather}
%
In this formulation, the pressures $p_{\|}(\psi,B)$ and $p_{\perp}(\psi,B)$ are functions of two variables: the magnetic flux $\psi$ and the magnetic field modulus $B$. The equation becomes essentially nonlinear, since the magnetic field modulus
\begin{gather}
\label{01.02:006}
B = \sqrt{|\vnabla \psi|^{2} + \mu_{0}^{2}F^{2}/\sigma^{2}}/R 
\end{gather}
%
itself contains the derivatives of the desired function $\psi$.

%
For equilibrium to exist, the condition of longitudinal equilibrium must be satisfied
\begin{gather}
\label{01.02:007}
\left( 
    \pdv{p_{\|}}{B} 
\right)_{\psi} 
= 
\frac{p_{\|} - p_{\perp}}{B}
,
\end{gather}
%
which relates the functions $p_{\|}(\psi,B)$ and $p_{\perp}(\psi,B)$, allowing $p_{\|}(\psi,B)$ to be expressed in terms of $p_{\perp}(\psi,B)$ or $p_{\perp}(\psi,B)$ in terms of $p_{\|}(\psi,B)$.

%
In Grad's article \cite{Grad1967PF_10_137}, equation \eqref{01.02:002} is numbered (5.12). Grad points out that the corresponding boundary value problem is well-posed only if two inequalities
\begin{gather}
\label{01.02:008}
\sigma >0
,
\\
\label{01.02:009}
\pdv{}{B}\left( p_{\bot}+B^{2}/2\mu_{0} \right)
>0
\end{gather}
are satisfied everywhere. Nowadays, these conditions are interpreted as a guarantee of the absence of the firehose and mirror instability, respectively. Grad also notes that the current density vector 
\begin{gather}
\label{01.02:010}
\vec{J}
=
\frac{1}{\mu_{0}}
\vnabla\times\vec{B}
\end{gather}
does not lie on the surface $\psi$ in the anisotropic problem, but
\begin{gather}
\label{01.02:011}
\vec{K}
=
\frac{1}{\mu_{0}}
\vnabla\times\left(
    \sigma\, \vec{B}
\right)
\end{gather}
takes its place. The derivation of the generalized Grad--Shafranov equation is given in Appendix \ref{A2}. It is also proven there that the modified current flux $F=F(\psi)$, defined by Eq.~\eqref{01.02:005} is a surface function, i.e.\ it depends on $\psi$, but not on $\psi$ and $B$, despite the fact that the current flux $i_{p}$, as well as the anisotropy parameter $\sigma$, are not surface functions

\medskip

Mathematical properties of the Grad--Shafranov and Mercier--Cotsaftis equations have attracted the attention of professional mathematicians. Several papers, such as \cite{Kostomarov+2008MathModelling_20_1, JeyakumarPfefferleHoleQu2021JPP_87_905870506}, have provided examples of correctly posed boundary value problems and established conditions for existence and uniqueness of solutions to the Grad--Shafranov equation both in the standard isotropic case and when accounting for pressure anisotropy and toroidal flow.

%
In toroidal systems, pressure anisotropy significantly affects the magnitude of the radial displacement of the plasma column in a tokamak, which was first calculated in the isotropic plasma approximation by V.~Shafranov \cite{Shafranov1962AE_13_521, Shafranov1966RevPlasPhys_2_103} and is therefore named after him. Anisotropy can either increase or decrease the Shafranov shift compared to the isotropic case, depending on the sign of the difference $p_{\bot}-p_{\|}$ \cite{MaddenHastie1994NF_34_519, Iacono+1990PFB_2_1794, LepikhinPustovitov2012ECA_36F_P5029, Pustovitov2013PPR_39_605}.

The global index contains approximately 800--1,200 articles in scientific journals citing GGSE. These include articles in leading journals, some of which are represented in Table \ref{tab:journals}, that explore extensions of the classical Grad--Shafranov equation to cases involving plasma flows, anisotropic pressure, or non-axially symmetric configurations.

The term GGSE is rarely used in the title of a book, but there are approximately 15-30 specialized monographs where the generalized Grad--Shafranov equation is discussed in detail in chapters devoted to magnetohydrodynamics and tokamak physics. The key authors here are V.~Shafranov, H.~Grad, as well as modern researchers such as G.~Throumoulopoulos \cite{ThroumoulopoulosPantis1986NF_26_1501, KuiroukidisThroumoulopoulos2016PoP_23_112508, KaltsasThroumoulopoulos2016PhyLettA_380_3373,  KuiroukidisKaltsasThroumoulopoulos2024PoP_31_042503}.

\subsection{Mercier--Cotsaftis equation}
\label{s01.03}

In its modern formulation, the Mercier--Cotsaftis equation is an alternative form 
\begin{gather}
\label{01.03:001}
\vnabla \cdot \left( \frac{\sigma}{R^{2}} \vnabla \psi \right) 
= 
- 
\mu_{0} \pdv{p_{\|}(\psi, B)}{\psi} 
- 
\mu_{0}^{2} \frac{F(\psi)}{\sigma R^{2}} \odv{F}{\psi}
\end{gather}
of  equation (5.42) in Grad's paper \cite{Grad1967PF_10_137}. The main difference from its replica \eqref{01.02:002} is the modification of the Shafranov operator on the left-hand side by combining it with the first term on the right-hand side.

%
Equation \eqref{01.03:001} preserves the elliptic type only for $\sigma > 0$. Violation of this condition leads to the development of firehose instability. Hence the Mercier--Cotsaftis equation allows one to take into account the effects of ``mirror'' and ``firehose'' instabilities directly from the equilibrium structure. The solvability condition of equation \eqref{01.03:001} is closely related to the fulfillment of the Mercier criterion for local stability of the plasma column \cite{Mercier1960CompRendAcadSci_250_1010, Mercier1961NF_1_47}. Thus, this is not just an alternative name, but an extension of the theory to systems where thermodynamic equilibrium is not scalar.

%
In the original version, the Mercier--Cotsaftis equation does not have much in common with its modern formulation \eqref{01.03:001}; even the notation of physical quantities used in \cite{MercierCotsaftis1961CompRendAcadSci_252_2203, MercierCotsaftis1961NF_1_121} may seem strange. In the short communication to the French Academy of Sciences \cite{MercierCotsaftis1961CompRendAcadSci_252_2203}, which, according to French academic tradition, preceded the publication of a detailed article \cite{MercierCotsaftis1961NF_1_121} to quickly establish the priority of the discovery, the equilibrium equation for anisotropic plasma is completely absent. Apparently, the authors were primarily interested in issues of plasma stability in toroidal magnetic confinement systems. In the final article \cite{MercierCotsaftis1961NF_1_121}, published in French, equation (8) is similar to the modern form of MCE \eqref{01.03:001}, but instead of the property $F=F(\psi)=\sigma\,i_{p}$, proved by H.~Grad, a more restrictive hypothesis $\sigma=\sigma(\psi)$ and $i_{p}=i_{p}(\psi)$ was used, which even migrated to some recent articles \cite{Clemente1993NF_33_963, SouzaViana2019PoP_26_042502, SouzaViana2020BrazJPhys_50_136}.

Most publications mentioning the Mercier--Cotsaftis equation date back to the 1960s--1980s; this was a period of active theoretical development of MHD stability without excessive computer modeling. In modern works, this equation is more often cited as the classical basis for analyzing Mercier criteria \cite{Mercier1960CompRendAcadSci_250_1010, Mercier1961NF_1_47}. However, since 2015, the work of Mercier and Cotsaftis has again attracted the attention of researchers. It is also worth noting that developers of numerical codes (e.g., W.~Zwingmann, L.-G.~Ericsson, and P.~Stubberfield \cite{ZwingmannErikssonStubberfield2001PPCF_43_1441}) cite both Grad's 1967 paper and Mercier and Cotsaftis's 1961 paper.

\section{Equilibrium equation for mirror traps}
\label{s03}

\begin{figure}
    \includegraphics[width=0.49\textwidth]{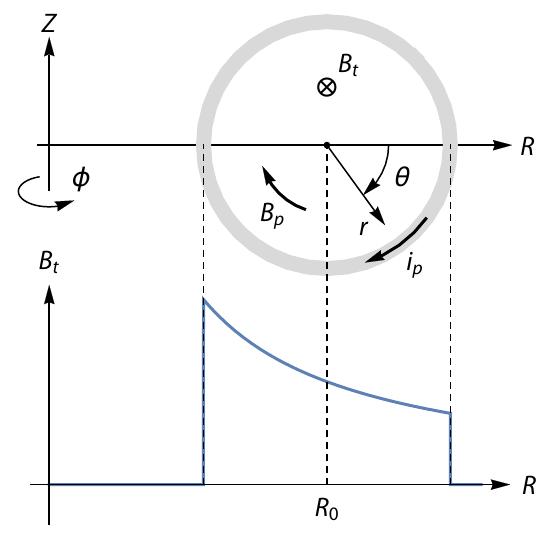}
    \caption{
%
In a symmetric torus, the toroidal magnetic field $B_{t}$ decreases inversely proportional to the distance $R$ from the $Z$ axis, since according to Stokes's theorem $2\pi R B_{t}(R) = \mu_{0}I_{p}$, where $I_{p}=2\pi i_{p}$ is the poloidal current flowing through a circle of radius $R$. Inside the torus, $I_{p}=IN=\const$, where $I$ is the current in the coil wound on the torus, and $N$ is the number of coils. Outside the torus, \mbox{$I_{p}=0$}.
In the theory of mirror traps, a cylindrical coordinate system $\{r,\theta ,z\}$ is used (where $z\approx R_{0}\phi$ and $R_{0}\to\infty$), see equation~\eqref{00:00}; or $\{R,\phi,Z\}$ with $R_{0}=0$; see problem 24.5 in \cite{Kotelnikov2025V2e}.
    }
\label{fig:208-02}
\end{figure}
%
The application of the Grad--Shafranov equation to axially symmetric open traps such as GDT \cite{IvanovPrikhodko2013PPCF_55_063001}, GDMT \cite{Skovorodin+2023PPR_49_1039} or WHAM \cite{WHAM2020} is greatly simplified since such systems lack a poloidal current, $i_{p}=0$, so that the Mercier--Cotsaftis equation reduces to
\begin{gather}
\label{01.04:001}
\vnabla \cdot \left( \frac{\sigma}{R^{2}} \vnabla \psi \right) 
= 
- 
\mu_{0} \pdv{p_{\|}(\psi, B)}{\psi} 
.
\end{gather}
Examples of analytical solutions of the classical Grad--Shafranov equation describing equilibrium in such traps can be found in modern textbooks, for example, \cite{Kotelnikov2025V2e}; see Fig.~\ref{fig:208-02}. However, until recently, neither the classical Grad--Shafranov equation nor its adaptations for anisotropic plasmas have been encountered in journal articles devoted to mirror traps. The point is that for the above-mentioned ``long and thin'' traps, sufficient accuracy is ensured by the so-called paraxial approximation, when the original equation of equilibrium of anisotropic plasma \eqref{01.02:001} is simplified to a local balance of transverse pressure
\begin{equation}
\label{01.04:002}
    p_{\perp} + \frac{B^{2}}{2\mu_{0}} 
    = 
    \frac{B_{v}^{2}}{2\mu_{0}}
    ,
\end{equation}
%
where $B_{v}$ is the absolute value of the magnetic field on the trap axis in the absence of plasma. Such an equilibrium is physically realizable if the macroscopic stability criteria \eqref{01.02:009} and \eqref{01.02:010} are met.

%
The situation changed with the advent of the idea of a diamagnetic trap \cite{Beklemishev2016PoP_23_082506}. As plasma accumulates in such a trap, a bubble is inflated from which the magnetic field is almost completely expelled. At the ends of the bubble, the curvature of the magnetic field lines increases so much that the paraxial approximation ceases to work, therefore, a more accurate description of the equilibrium configuration of the plasma is needed. The Grad--Shafranov equation \eqref{01.01:008} or its modification \eqref{01.02:002}, \eqref{01.03:001} for anisotropic plasma could serve as a tool for such a description, if not for the circumstance pointed out in section \ref{s01}: the right-hand side of the equation \eqref{00:00} cannot be represented as a function of only the magnetic flux and its derivatives. The equation can continue to be called the Grad--Shafranov equation, as the authors of the cited papers do \cite{KhristoBeklemishev2025JPP_91_E3, Chernoshtanov2025arXiv2512_01780}, but a grumpy mathematician would say that this equation is of a completely different type: integro-differential, not differential. Who knows, maybe one day the equation \eqref{00:00} will be called the Beklemishev--Khristo equation?

%
Nevertheless, I would venture to suggest that the equation \eqref{01.04:001} is quite suitable for describing intermediate states of equilibrium in a diamagnetic trap, when the magnetic field has not yet been completely expelled from the diamagnetic bubble.

\section{Afterword}

%
In the history of science, it sometimes happens that a phenomenon is named not by its discoverer, but by a popularizer or colleague, and sometimes the name of the discoverer is completely replaced by that of someone who ``explained it better.'' The story of the Mercier--Cotsaftis equation shows just such a case. Harold Grad proposed a clearer derivation of the Mercier--Cotsaftis equation than its discoverers. Grad's greater authority and greater fame in the scientific community also played a role. But the history of science also knows of examples where, years later, justice was at least partially restored.

%
This is what happened with the Hubble constant. Although the law of the expansion of the Universe bears Hubble's name, it was first mathematically derived and theoretically substantiated by the Belgian priest and astronomer Georges Lema\^{i}tre in 1927 (two years before Hubble's publication). Hubble, in turn, provided observational evidence.

Result: It was only in 2018 that the International Astronomical Union recommended renaming the law to the ``Hubble-Lema\^{i}tre law.''

\medskip

%
Considering the multiplicity of variants of the name of the equilibrium equation of anisotropic plasma with the inclusion of the words ``Grad--Shafranov'', in addition to restoring historical justice, it would even be convenient to get rid of this confusion and name this equation after Claude Mercier and Michel Cotsaftis.

\appendix

\section{Derivation of the Grad--Shafranov equation}
\label{A1}

%
The derivation of the canonical Grad--Shafranov equation is based on a combined solution of the equations of ideal magnetohydrodynamics (MHD) and Maxwell's equations under the assumption of axial symmetry ($\partial/\partial\phi = 0$) in cylindrical coordinates $(R, \phi, Z)$. Due to the absence of divergence
\begin{gather}
\label{A1:001}
\vnabla \cdot \vec{B} = 0
,
\end{gather}
the magnetic field $\vec{B}$ can be expressed in terms of the poloidal magnetic flux $\psi(R, Z)$ and the poloidal current flux $i_{p}(R, Z)$ using the formula \eqref{01.01:006}:
\begin{gather}
\label{A1:002}
\vec{B} 
= 
\frac{1}{R} \left( \vnabla \psi \times \vec{e}_{\phi} \right) 
+ 
\mu_{0}\frac{i_{p}}{R} \vec{e}_{\phi}
.
\end{gather}
Applying Ampere's law
\begin{gather}
\label{A1:003}
\mu_{0} \vec{J}
=
\vnabla \times \vec{B}
\end{gather}
and taking into account the properties of the rotor in cylindrical symmetry, the current density $\vec{J}$ takes the form
\begin{gather}
\label{A1:004}
\vec{J} 
= 
\frac{1}{R} \odv{i_{p}}{\psi} \vnabla \psi \times \vec{e}_{\phi} 
- 
\frac{1}{\mu_{0}R} \Delta^* \psi\,\vec{e}_{\phi},
\end{gather}
where 
\begin{gather}
\label{A1:005}
\Delta^* \psi 
= 
R \pdv{}{R} 
\left( \frac{1}{R} \pdv{\psi}{R} \right) 
+ 
\pdv[2]{\psi}{Z}
,
\end{gather}
is the Shafranov operator \eqref{01.01:003}, and $\vec{e}_{\phi}=\vnabla\phi/|\vnabla\phi|$ is the unit vector in the direction of the angle $\phi$.

%
The equilibrium state of an isotropic plasma is described by the force balance equation \eqref{01.01:011}:
\begin{gather}
\label{A1:006}
\vnabla p = \vec{J} \times \vec{B}
.
\end{gather}
From the scalar product \eqref{A1:006} with the vectors $\vec{B}$ and $\vec{J}$ it follows that $(\vec{B} \cdot \vnabla) p = 0$ and $(\vec{J} \cdot \vnabla) p = 0$. This means that $p$ and $i_{p}$ are surface functions, that is, they depend only on the magnetic flux: 
\begin{gather}
\label{A1:007}
p = p(\psi)
,
\qquad 
i_{p} = i_{p}(\psi)
. 
\end{gather}
Thus, $\vnabla p = p'(\psi) \vnabla \psi$.
%
%
Substituting the expressions \eqref{A1:002} and \eqref{A1:004} for $\vec{J}$ and $\vec{B}$ into equation \eqref{A1:006} and projecting it onto the $\vnabla \psi$ direction, we obtain the equation
\begin{gather}
\label{A1:008}
p'(\psi) \vnabla \psi 
= 
\left( 
    -
    \frac{1}{\mu_{0} R^{2}} \Delta^* \psi 
    - 
    \frac{\mu_{0}}{ R^{2}} i_{p} i_{p}'(\psi) 
\right) 
\vnabla \psi
,
\end{gather}
%
which, after cancellation by $\vnabla \psi$ and rearrangement of terms, takes the form of the classical Grad equation--Shafranov \eqref{01.01:008}:
\begin{gather}
\label{A1:009}
\Delta^* \psi 
= 
-\mu_{0} R^{2} \odv{p}{\psi} 
- \mu_{0}^{2}i_{p} \odv{i_{p}}{\psi}
.
\end{gather}
This is an elliptic equation with a nonlinear right-hand side. It determines the geometry of the magnetic field for given pressure $p(\psi)$ and current flux $i_{p}(\psi)$ profiles.

\section{Derivation of the Mercier--Cotsaftis equation}
\label{A2}

An anisotropic static equilibrium can be described by
the system of equations 
\begin{gather}
\label{A002:001}
\vnabla \cdot \mathbf{P} 
= 
\vec{J} \times \vec{B}
,\\
\label{A002:002}
\mathbf{P} 
= 
p_{\perp} \mathbf{I} 
+ 
\frac{p_{\|} - p_{\perp}}{B^{2}}
\vec{B} \vec{B}
,\\
\label{A002:003}
\vec{J}
=
\frac{1}{\mu_{0}}
\vnabla\times\vec{B}
,\\
\label{A002:004}
\vnabla \cdot \vec{B}
=
0
,
\end{gather}
with $p_{\|}$ and $p_{\perp}$ being functions of the magnetic flux function $\psi$ and the magnitude of the magnetic field $B$. It is convenient to introduce the functions
\begin{gather}
\label{A002:005}
\sigma 
=
1
-
\mu_{0}
\frac{1}{B}\left(
    \pdv{p_{\|}}{B}
\right)_{\psi}
,\\
\label{A002:006}
\tau 
=
1
+
\mu_{0}
\frac{1}{B}\left(
    \pdv{p_{\perp}}{B}
\right)_{\psi}
,\\
\intertext{and}
\label{A002:007}
\vec{K}
=
\frac{1}{\mu_{0}}
\vnabla\times\left( \sigma\, \vec{B} \right)
.
\end{gather}
Hence Eq.~\eqref{A002:001} reduces to
\begin{gather}
\label{A002:008}
\vec{K} \times \vec{B}
=
\left(
    \pdv{p_{\|}}{\psi}
\right)_{B}
\vnabla\psi
\\
\intertext{and}
\label{A002:009}
\mu_{0}
\frac{1}{B}\left(
    \pdv{p_{\|}}{B}
\right)_{\psi}
=
1
-
\sigma 
=
\mu_{0}
\frac{p_{\|}-p_{\perp}}{B^{2}}
.
\end{gather}
The generalized pressure surfaces, which contain the two vector fields $\vec{B}$ and $\vec{K}$, have many properties analogous to constant pressure surfaces in the isotropic pressure case. The magnetic field can be written as
\begin{gather}
\label{A002:010}
\vec{B} 
= 
\vnabla \psi \times \vnabla \phi 
+ 
\mu_{0}{i_{p}} \vnabla \phi
,
\end{gather}
where 
\begin{gather}
\label{A002:010a}
\psi(R,Z)
=
\int_{0}^{R}
R'\, B_{Z}(R',Z)\odif{R'}
\end{gather}
is the reduced (i.e.\ divided by $2\pi$) poloidal magnetic flux between system symmetry axis and the magnetic surface passing through the specified point $\{R,Z\}$. The $\vnabla\phi$ component of Eq.~\eqref{A002:008} makes $\vec{K}\cdot \vnabla\psi=0$. The use of Eqs.~\eqref{A002:007} and~\eqref{A002:010} then shows that $\sigma \,i_{p}$
is a function only of $\psi$; so that we can define
\begin{gather}
\label{A002:011}
F(\psi)
=
\sigma \,i_{p}
.
\end{gather}
Taking the $\vnabla \psi$ components of Eq.~\eqref{A002:008} and using the relations
between $\vec{J}$, $\vec{B}$, and $\vec{K}$ in Eqs.~\eqref{A002:003} and~\eqref{A002:007}, we obtain 
the Mercier--Cotsaftis equation
\begin{gather}
\vnabla \cdot \left( \frac{\sigma}{R^{2}} \vnabla \psi \right) 
= 
- 
\mu_{0} \pdv{p_{\|}(\psi, B)}{\psi} 
- 
\mu_{0}^{2} \frac{F(\psi)}{\sigma R^{2}} \odv{F}{\psi}
.
\end{gather}

\typeout{*************************************** 2524}
\printbibliography[]

@ARTICLE{KhristoBeklemishev2025JPP_91_E3,
  AUTHOR = {Khristo, M. S. and Beklemishev, A. D.},
  DATE = {2025},
  DOI = {10.1017/S0022377824001417},
  JOURNALTITLE = {Journal of Plasma Physics},
  NUMBER = {1},
  PAGES = {E3},
  TITLE = {Plasma equilibrium in diamagnetic trap with neutral beam injection},
  VOLUME = {91},
}

@ARTICLE{Beklemishev2016PoP_23_082506,
  AUTHOR = {Beklemishev, A. D.},
  URL = {https://doi.org/10.1063/1.4960129},
  DATE = {2016},
  DOI = {10.1063/1.4960129},
  JOURNALTITLE = {Physics of Plasmas},
  NUMBER = {8},
  PAGES = {082506},
  TITLE = {Diamagnetic “bubble” equilibria in linear traps},
  VOLUME = {23},
}

@MISC{Chernoshtanov2025arXiv2512_01780,
  AUTHOR = {Chernoshtanov, I.},
  LANGUAGE = {english},
  URL = {https://arxiv.org/abs/2512.01780},
  DATE = {2025},
  EPRINT = {2512.01780},
  EPRINTCLASS = {physics.plasm-ph},
  EPRINTTYPE = {arXiv},
  LANGID = {english},
  TITLE = {High-beta equilibrium in mirror machine with population of fast sloshing ions},
}

@ARTICLE{GradRubin1958JNuclEnergy_7_284,
  AUTHOR = {Grad, H. and Rubin, H.},
  PUBLISHER = {Pergamon},
  URL = {https://www.sciencedirect.com/science/article/pii/0891391958901396},
  DATE = {1958-09},
  DOI = {https://doi.org/10.1016/0891-3919(58)90139-6},
  ISSN = {0891-3919},
  JOURNALTITLE = {Journal of Nuclear Energy (1954)},
  NUMBER = {3},
  PAGES = {284--285},
  TITLE = {Hydromagnetic equilibria and force-free fields},
  VOLUME = {7},
}

@INPROCEEDINGS{GradRubin1958UN2_31_190,
  AUTHOR = {Grad, H. and Rubin, H.},
  LOCATION = {Geneva, Switzerland},
  ORGANIZATION = {United Nations},
  BOOKTITLE = {{}, Proceedings of the Second United Nations Conference on the Peaceful Uses of Atomic Energy (September 1–13, 1958)},
  DATE = {1958-09},
  PAGES = {190--197},
  TITLE = {Hydromagnetic Equilibria and Force-Free Fields},
  VOLUME = {31},
}

@ARTICLE{Shafranov1957JETPh_33_710,
  AUTHOR = {Шафранов, В. Д.},
  LANGUAGE = {russian},
  DATE = {1957-09},
  JOURNALTITLE = {Журнал экспериментальной и теоретической физики},
  LANGID = {russian},
  NUMBER = {3},
  PAGES = {710--722},
  TITLE = {О равновесных магнитогидродинамических конфигурациях},
  VOLUME = {33},
}

@ARTICLE{Shafranov1958SovPhysJETP_6_545,
  AUTHOR = {Shafranov, V. D.},
  LANGUAGE = {english},
  DATE = {1958-03},
  JOURNALTITLE = {Sov. Phys. JETP},
  LANGID = {english},
  PAGES = {545},
  TITLE = {On equilibrium magnetohydrodynamic configurations},
  VOLUME = {6},
}

@ARTICLE{Shafranov1960,
  AUTHOR = {Shafranov, V. D.},
  DATE = {1960},
  JOURNALTITLE = {Soviet Physics JETP},
  NUMBER = {4},
  PAGES = {775--781},
  TITLE = {Equilibrium of a Plasma Toroid in a Magnetic Field},
  VOLUME = {10},
}

@ARTICLE{MercierCotsaftis1961CompRendAcadSci_252_2203,
  AUTHOR = {Mercier, C. and Cotsaftis, M.},
  LANGUAGE = {french},
  DATE = {1961-03},
  JOURNALTITLE = {Comptes rendus hebdomadaires des séances de l'Académie des sciences},
  LANGID = {french},
  PAGES = {2203--2205},
  TITLE = {Équilibre et stabilité pour les systèmes toroïdaux magnétohydrodynamiques en pression scalaire au voisinage d'un axe magnétique},
  VOLUME = {252},
}

@ARTICLE{MercierCotsaftis1961NF_1_121,
  AUTHOR = {Mercier, C. and Cotsaftis, M.},
  LANGUAGE = {french},
  URL = {https://dx.doi.org/10.1088/0029-5515/1/2/005},
  DATE = {1961-03},
  DOI = {10.1088/0029-5515/1/2/004},
  JOURNALTITLE = {Nuclear Fusion},
  LANGID = {english},
  NUMBER = {2},
  PAGES = {121--145},
  TITLE = {Equilibre et Stabilite d'un Plasma en Symetrie de Revolution avec Pression Anisotrope},
  VOLUME = {1},
}

@INPROCEEDINGS{Grad1967ProcSympApplMath_18_162,
  AUTHOR = {Grad, H.},
  LOCATION = {Providence, RI},
  PUBLISHER = {American Mathematical Society},
  BOOKTITLE = {Magneto-Fluid and Plasma Dynamics},
  DATE = {1967},
  DOI = {10.1090/psapm/018},
  ISBN = {978-0-8218-1318-8},
  PAGES = {162--182},
  SERIES = {Proceedings of Symposia in Applied Mathematics},
  TITLE = {Magneto-fluid and plasma dynamics},
  VOLUME = {18},
}

@ARTICLE{Grad1967PF_10_137,
  AUTHOR = {Grad, H.},
  PUBLISHER = {American Mathematical Society},
  DATE = {1967},
  DOI = {10.1063/1.1761965},
  JOURNALTITLE = {Physics of Fluids},
  NUMBER = {1},
  PAGES = {137--154},
  TITLE = {Toroidal Containment of a Plasma},
  VOLUME = {10},
}

@ARTICLE{Kostomarov+2008MathModelling_20_1,
  AUTHOR = {Костомаров, Д. П. and Медведев, С. Ю. and Сычугов, Д. Ю.},
  LANGUAGE = {russian},
  PUBLISHER = {Федеральное государственное бюджетное учреждение" Российская академия наук"},
  DATE = {2008},
  JOURNALTITLE = {Математическое моделирование},
  LANGID = {russian},
  NUMBER = {5},
  PAGES = {1--35},
  TITLE = {Математическое моделирование МГД равновесия плазмы},
  VOLUME = {20},
}

@ARTICLE{HamFarrell2024NF_64_034001,
  AUTHOR = {Ham, C. J. and Farrell, P. E.},
  PUBLISHER = {IOP Publishing},
  URL = {https://doi.org/10.1088/1741-4326/ad1d77},
  DATE = {2024-02},
  DOI = {10.1088/1741-4326/ad1d77},
  JOURNALTITLE = {Nuclear Fusion},
  NUMBER = {3},
  PAGES = {034001},
  TITLE = {On multiple solutions of the Grad–Shafranov equation},
  VOLUME = {64},
}

@ARTICLE{JeyakumarPfefferleHoleQu2021JPP_87_905870506,
  AUTHOR = {Jeyakumar, S. and Pfefferlé, D. and Hole, M.J. and Qu, Z.S.},
  DATE = {2021},
  DOI = {10.1017/S002237782100088X},
  JOURNALTITLE = {Journal of Plasma Physics},
  NUMBER = {5},
  PAGES = {905870506},
  TITLE = {Analysis of the isotropic and anisotropic Grad–Shafranov equation},
  VOLUME = {87},
}

@ARTICLE{Shafranov1962AE_13_521,
  AUTHOR = {Шафранов, В. Д.},
  LANGUAGE = {russian},
  URL = {elib.biblioatom.ru/text/atomnaya-energiya_t13-6_1962/0000/},
  DATE = {1962},
  JOURNALTITLE = {Атомная энергия},
  LANGID = {russian},
  NUMBER = {6},
  PAGES = {521--529},
  TITLE = {Равновесие тороидального плазменного шнура в магнитном поле},
  VOLUME = {13},
}

@ARTICLE{Shafranov1966RevPlasPhys_2_103,
  AUTHOR = {Shafranov, V. D.},
  PUBLISHER = {Consultants Bureau},
  DATE = {1966},
  JOURNALTITLE = {Reviews of Plasma Physics},
  PAGES = {103},
  TITLE = {Plasma equilibrium in a magnetic field},
  VOLUME = {2},
}

@ARTICLE{MaddenHastie1994NF_34_519,
  AUTHOR = {Madden, N. A. and Hastie, R. J.},
  URL = {https://doi.org/10.1088/0029-5515/34/4/I05},
  DATE = {1994-04},
  DOI = {10.1088/0029-5515/34/4/I05},
  JOURNALTITLE = {Nuclear Fusion},
  NUMBER = {4},
  PAGES = {519},
  TITLE = {Tokamak equilibrium with anisotropic pressure},
  VOLUME = {34},
}

@ARTICLE{Iacono+1990PFB_2_1794,
  AUTHOR = {Iacono, R. and Bondeson, A. and Troyon, F. and Gruber, R.},
  LANGUAGE = {english},
  URL = {https://doi.org/10.1063/1.859451},
  DATE = {1990},
  DOI = {10.1063/1.859451},
  EPRINT = {https://pubs.aip.org/aip/pfb/article-pdf/2/8/1794/12324223/1794_1_online.pdf},
  ISSN = {0899-8221},
  JOURNALTITLE = {Physics of Fluids B: Plasma Physics},
  LANGID = {english},
  NUMBER = {8},
  PAGES = {1794--1803},
  TITLE = {Axisymmetric toroidal equilibrium with flow and anisotropic pressure},
  VOLUME = {2},
}

@INPROCEEDINGS{LepikhinPustovitov2012ECA_36F_P5029,
  AUTHOR = {Lepikhin, N. D. and Pustovitov, V. D.},
  URL = {http://info.fusion.ciemat.es},
  BOOKTITLE = {39th EPS Conference on Plasma Physics},
  DATE = {2012},
  PAGES = {P5.029},
  SERIES = {Europhysics Conference Abstracts},
  TITLE = {Shafranov shift at strong plasma anisotropy in tokamaks and stellarators},
  VOLUME = {36F},
}

@ARTICLE{Pustovitov2013PPR_39_605,
  AUTHOR = {Lepikhin, N. D. and Pustovitov, V. D.},
  PUBLISHER = {Springer},
  DATE = {2013},
  DOI = {10.1134/S1063780X13080059},
  JOURNALTITLE = {Plasma Physics Reports},
  NUMBER = {8},
  PAGES = {605--614},
  TITLE = {Analytical Modeling of Equilibrium of Strongly Anisotropic Plasma in Tokamaks and Stellarators},
  VOLUME = {39},
}

@ARTICLE{ThroumoulopoulosPantis1986NF_26_1501,
  AUTHOR = {Throumoulopoulos, G. N. and Pantis, G.},
  URL = {https://doi.org/10.1088/0029-5515/26/11/005},
  DATE = {1986-11},
  DOI = {10.1088/0029-5515/26/11/005},
  JOURNALTITLE = {Nuclear Fusion},
  NUMBER = {11},
  PAGES = {1501},
  TITLE = {The Grad-Shafranov equation under a conformal mapping transformation: Analytic solutions with emphasis on compact toroidal configurations},
  VOLUME = {26},
}

@ARTICLE{KuiroukidisThroumoulopoulos2016PoP_23_112508,
  AUTHOR = {Kuiroukidis, A. I. and Throumoulopoulos, G. N.},
  URL = {https://doi.org/10.1063/1.4968235},
  DATE = {2016-11},
  DOI = {10.1063/1.4968235},
  EPRINT = {https://pubs.aip.org/aip/pop/article-pdf/doi/10.1063/1.4968235/16068901/112508_1_online.pdf},
  ISSN = {1070-664X},
  JOURNALTITLE = {Physics of Plasmas},
  NUMBER = {11},
  PAGES = {112508},
  TITLE = {New classes of exact solutions to the Grad-Shafranov equation with arbitrary flow using Lie-point symmetries},
  VOLUME = {23},
}

@ARTICLE{KaltsasThroumoulopoulos2016PhyLettA_380_3373,
  AUTHOR = {Kaltsas, D. A. and Throumoulopoulos, G. N.},
  URL = {https://www.sciencedirect.com/science/article/pii/S0375960116305527},
  DATE = {2016},
  DOI = {https://doi.org/10.1016/j.physleta.2016.08.011},
  ISSN = {0375-9601},
  JOURNALTITLE = {Physics Letters A},
  NUMBER = {41},
  PAGES = {3373--3377},
  TITLE = {Exact solutions of the Grad–Shafranov equation via similarity reduction and applications to magnetically confined plasmas},
  VOLUME = {380},
}

@ARTICLE{KuiroukidisKaltsasThroumoulopoulos2024PoP_31_042503,
  AUTHOR = {Kuiroukidis, A. I. and Kaltsas, D. A. and Throumoulopoulos, G. N.},
  URL = {https://doi.org/10.1063/5.0198558},
  DATE = {2024-04},
  DOI = {10.1063/5.0198558},
  EPRINT = {https://pubs.aip.org/aip/pop/article-pdf/doi/10.1063/5.0198558/19870227/042503_1_5.0198558.pdf},
  ISSN = {1070-664X},
  JOURNALTITLE = {Physics of Plasmas},
  NUMBER = {4},
  PAGES = {042503},
  TITLE = {A similarity reduction of the generalized Grad–Shafranov equation},
  VOLUME = {31},
}

@ARTICLE{Mercier1960CompRendAcadSci_250_1010,
  AUTHOR = {Mercier, C.},
  LANGUAGE = {french},
  DATE = {1960},
  JOURNALTITLE = {Comptes Rendus de l'Académie des Sciences},
  PAGES = {1010--1012},
  TITLE = {Un critère nécessaire de stabilité hydromagnétique pour un plasma en symétrie de révolution},
  VOLUME = {250},
}

@ARTICLE{Mercier1961NF_1_47,
  AUTHOR = {Mercier, C.},
  LANGUAGE = {french},
  PUBLISHER = {IAEA},
  DATE = {1961},
  DOI = {10.1088/0029-5515/1/1/005},
  JOURNALTITLE = {Nuclear Fusion},
  NUMBER = {1},
  PAGES = {47--53},
  TITLE = {Critère de stabilité d'un système toroïdal hydromagnétique en pression scalaire},
  VOLUME = {1},
}

@ARTICLE{Clemente1993NF_33_963,
  AUTHOR = {Clemente, R. A.},
  URL = {https://doi.org/10.1088/0029-5515/33/6/I12},
  DATE = {1993-06},
  DOI = {10.1088/0029-5515/33/6/I12},
  JOURNALTITLE = {Nuclear Fusion},
  NUMBER = {6},
  PAGES = {963},
  TITLE = {Anisotropic axisymmetric equilibria via an analytic method},
  VOLUME = {33},
}

@ARTICLE{SouzaViana2019PoP_26_042502,
  AUTHOR = {Souza, L. C. and Viana, R. L.},
  URL = {https://doi.org/10.1063/1.5084793},
  DATE = {2019-04},
  DOI = {10.1063/1.5084793},
  EPRINT = {https://pubs.aip.org/aip/pop/article-pdf/doi/10.1063/1.5084793/14020274/042502_1_online.pdf},
  ISSN = {1070-664X},
  JOURNALTITLE = {Physics of Plasmas},
  NUMBER = {4},
  PAGES = {042502},
  TITLE = {Anisotropic MHD equilibria in symmetric systems},
  VOLUME = {26},
}

@ARTICLE{SouzaViana2020BrazJPhys_50_136,
  AUTHOR = {Souza, L. C. and Viana, R. L.},
  PUBLISHER = {Springer},
  DATE = {2020},
  JOURNALTITLE = {Brazilian Journal of Physics},
  NUMBER = {2},
  PAGES = {136--142},
  TITLE = {Anisotropic Axisymmetric MHD Equilibria in Spheroidal Coordinates},
  VOLUME = {50},
}

@ARTICLE{ZwingmannErikssonStubberfield2001PPCF_43_1441,
  AUTHOR = {Zwingmann, W. and Eriksson, L.-G. and Stubberfield, P.},
  LANGUAGE = {english},
  URL = {https://doi.org/10.1088/0741-3335/43/11/302},
  DATE = {2001-10},
  DOI = {10.1088/0741-3335/43/11/302},
  JOURNALTITLE = {Plasma Physics and Controlled Fusion},
  LANGID = {english},
  NUMBER = {11},
  PAGES = {1441},
  TITLE = {Equilibrium analysis of tokamak discharges with anisotropic pressure},
  VOLUME = {43},
}

@BOOK{Kotelnikov2025V2e,
  AUTHOR = {Kotelnikov, I. A.},
  LANGUAGE = {english},
  LOCATION = {Saint Petersburg},
  PUBLISHER = {Lan'},
  DATE = {2025},
  EDITION = {5},
  ISBN = {978-5-507-50489-3},
  LANGID = {english},
  NOTE = {(in Russian)},
  PAGETOTAL = {448},
  SERIES = {Lectures on plasma physics},
  TITLE = {Magnetic hydrodynamics},
  VOLUME = {2},
}

@ARTICLE{IvanovPrikhodko2013PPCF_55_063001,
  AUTHOR = {Ivanov, A. A. and Prikhodko, V. V.},
  URL = {http://stacks.iop.org/0741-3335/55/i=6/a=063001},
  DATE = {2013},
  JOURNALTITLE = {Plasma Physics and Controlled Fusion},
  NUMBER = {6},
  PAGES = {063001},
  TITLE = {Gas-dynamic trap: an overview of the concept and experimental results},
  VOLUME = {55},
}

@ARTICLE{Skovorodin+2023PPR_49_1039,
  AUTHOR = {Skovorodin, D. I. and Chernoshtanov, I. S. and Amirov, V. Kh. and Astrelin, V. T. and Bagryanskii, P. A. and Beklemishev, A. D. and Burdakov, A. V. and Gorbovskii, A. I. and Kotel’nikov, I. A. and Magommedov, E. M. and Polosatkin, S. V. and Postupaev, V. V. and Prikhod’ko, V. V. and Savkin, V. Ya. and Soldatkina, E. I. and Solomakhin, A. L. and Sorokin, A. V. and Sudnikov, A. V. and Khristo, M. S. and Shiyankov, S. V. and Yakovlev, D. V. and Shcherbakov, V. I.},
  LANGUAGE = {english},
  URL = {https://doi.org/10.1134/S1063780X23600986},
  DATE = {2023},
  DOI = {10.1134/S1063780X23600986},
  JOURNALTITLE = {Plasma Physics Reports},
  LANGID = {english},
  NUMBER = {9},
  PAGES = {1039--1086},
  TITLE = {{G}as-{D}ynamic {M}ultiple-{M}irror {T}rap {GDMT}},
  VOLUME = {49},
}

@MISC{WHAM2020,
  AUTHOR = {Anderson, J. and Clark, M. and Forest, C. and Geiger, B. and Mirnov, V. and Oliva, S. and Pizzo, J. and Schmitz, O. and Wallace, J. and Kristofek, G. and Mumgaard, R. and Peterson, E. and Ram, A. and Whyte, D. and Wright, J. and Wukitch, S. and Green, D. and Harvey, R. and Petrov, Yu. V. and Srinivisan, B. and Hakim, A.},
  URL = {https://ui.adsabs.harvard.edu/abs/2020APS..DPPC20003A/abstract},
  DATE = {2020},
  HOWPUBLISHED = {APS Division of Plasma Physics Meeting 2020, abstract id.CP20.003},
  TITLE = {Introducing the {W}isconsin {HTS} {A}xisymmetric {M}irror},
}
\typeout{*************************************** 2526}%
\typeout{*************************************** 2527}%
\ErrorsOff*

\end{document}